\documentclass[paper]{revtex4}

\usepackage{graphicx}
\usepackage{epsfig}
\usepackage{fancyhdr}
\usepackage{empheq}
\usepackage{mathbbol}
\usepackage{wasysym}
\usepackage{pstricks}
\usepackage{bbm}
\usepackage{bm}
\usepackage{color}
\newcommand{\be}{\begin{equation}}
\newcommand{\ee}{\end{equation}}
\newcommand{\bea}{\begin{eqnarray}}
\newcommand{\eea}{\end{eqnarray}}

\begin{document}

\title{Heavier Higgs Particles:  Indications from Minimal Supersymmetry}

%

\author{L Maiani$^{*}$, AD Polosa$^{*}$} 
\author{V Riquer$^\dag$}
\affiliation{$^*$Dipartimento di Fisica, Sapienza Universit\`a di Roma, Piazzale A Moro 5, Roma, I-00185, Italy\\
$^\dag$Fondazione TERA, Via G Puccini 11, I-28100 Novara, Italy}

\begin{abstract}
We use the most recent data on the Higgs-like resonance $h$ observed at 125 GeV to  derive information about the mass  of the heavier Higgs particles predicted by Minimal Supersymmetry.  We treat as independent parameters the couplings of $h$ to top quark, beauty and massive vector bosons and, in this three dimensional space, we locate the point realizing the best fit to data and compare it to the position of the Standard Model point and to the region of coupling values accommodating heavier Higgs particles in Minimal Supersymmetry. 
We conclude that mass values $320\lesssim M_H\lesssim 360$~GeV are compatible at $2\sigma$ with the best fit of couplings to present data, larger values being  compatible at the $1\sigma$ level. Values of $1\lesssim \tan \beta\lesssim 6$ are compatible with data. \newline\newline
\end{abstract}

\maketitle

\thispagestyle{fancy}


In a recent paper~\cite{Maiani:2012ij}, we confronted Minimal Supersymmetry for the Higgs sector with the particle discovered by ATLAS and CMS at $125$~GeV. At that time, only the value of the mass was available, which, as noted also by other authors~\cite{Barger,Djuadi1}, is remarkably consistent with Minimal Supersymmetric Standard Model (MSSM), albeit with a value of the scalar top mass around $4$ TeV. The presence of a tenuous bump at about $320$~GeV in the CMS data for the $ZZ$ channel  could have been taken as a hint of the heavier, MSSM, $0^+$ Higgs particle, $H$.  With the higher statistics now available there is no trace left of such a structure.  

In this note, we confront MSSM with the most recent data of the resonance $h(125)$ provided by ATLAS~\cite{ATLAS}, CMS~\cite{CMS} and Tevatron~\cite{tevatron}. Mass is given as:
\be
M_h=\begin{pmatrix} 126.0\pm0.4\pm0.4,\;\mathrm{ATLAS} \\ 125.3\pm0.4\pm0.5,\;\;\;\;\;\mathrm{CMS}\end{pmatrix}
\label{massh}
\ee
For simplicity, we shall use in the computations of cross sections and decay rates the conventional value $M_h=125$ GeV.

The experimental signals for decays are given by the ratios:
\be
\mu_i= \frac{(\sigma_i\times \mathrm{BR}_i)_{\rm expt}}{(\sigma_i\times \mathrm{BR}_i)_{\rm SM}}
\label{muvalues}
\ee
for the channels $i= WW, ZZ, b \bar b, \tau \bar\tau, \gamma \gamma$, for ATLAS~\cite{ATLAS} and CMS~\cite{CMS}, and $i=b\bar b$ for Tevatron~\cite{tevatron}.
 
 We aim to obtain limits on the mass of $H$ and of its closeby partners $A$ and $H^{\pm}$ and on the $\tan\beta$ parameter.
 
  We find that, to one standard deviation, MSSM is consistent with present observation for $M_H\gtrsim 360$~GeV, and for $\tan\beta \gtrsim 1$.  

We study the problem in the $(c_t,c_b,c_W)$ space of  couplings of the $h$ resonance to the top and beauty quarks and to the vector bosons, $c_i$ representing the ratios of the MSSM to the Standard Model (SM) couplings.  We set $c_Z=c_W$ and use the same coupling $c_t$ for all up fermions, {\it e.g.} charm, and $c_b$ for all down fermions, {\it e.g.} $\tau$. 

Two different types of cross sections for the $h$ production appear into the definition of $\mu_i$ given above, namely the gluon fusion cross section $\sigma_{\rm gf}$ and the vector boson fusion one,  $\sigma_{_{\rm VBF}}$. These are parametrized according to
\bea
\label{sigmagf}
&&\sigma_{\rm gf}=(\sigma_{\rm gf})_{_{\rm SM}} \; c_{t}^2 \\
&&\sigma_{_{\rm VBF}}=(\sigma_{_{\rm  VBF}})_{_{\rm SM}} \; c_{W}^2 
\label{sigmavbf}
\eea
Cross section (\ref{sigmagf}) is dominated by the top quark loop and it applies to the first four decay channels listed above.
The cross section in (\ref{sigmavbf}) appears in the $\gamma\gamma$ channel due to the stringent cuts applied by ATLAS and CMS to the diphoton distribution so as to enhance the purity of the $\gamma\gamma$ signal. In this case we write:
\be
\bar \sigma =(\bar \sigma_{\rm gf})_{_{\rm SM}}\, c_t^2+(\bar \sigma_{_{\rm VBF}})_{_{\rm SM}}\, c_W^2
\ee
where the bar represents the effect of the experimental cuts. 
The ratio $\bar \sigma/\bar \sigma_{\rm SM}$ depends only on the ratio:
\be
R=\frac{\bar \sigma_{_{\rm VBF}}}{\bar \sigma_{\rm gf}}
\ee
In Ref. \cite{Maiani:2012ij} one finds a detailed calculation of $R$ at 7 TeV using ALPGEN~\cite{alpgen} supplemented by the cuts specified by CMS in Ref. \cite{Chatrchyan:2012dg}. No difference using ATLAS cuts is found. 
We found $R=2.97\;@\;7$ TeV, whereas we find, with the same tools,  $R=3.1\;@\; 8$ TeV. 
Neglecting such a small deviation, we can use the cross sections computed in  \cite{Maiani:2012ij}  to compare with the experimental values for $\mu_i$, which are obtained by combining data at 7 and 8 TeV.

Widths in Vector-Vector and fermion antifermion decays of $h$ are parametrized by rescaling the Standard Model values by the appropriate factors, $c_W^2$, etc. As for the $\gamma \gamma$ width, we write:
\be
\Gamma(\gamma \gamma)=\Gamma(\gamma \gamma)_{_{\rm SM}}\; \frac{|g_1 c_W+g_{1/2}c_t|^2}{|g_1 +g_{1/2}|^2}
\label{2gamma}
\ee
where $g_{1,1/2}$ are the loop contributions of $W$ and top\; \cite{marciano}, with:
\be
g_{12} = 8.32,\; g_{1/2}=-1.84
\ee

The total width is obtained from the sum of the individual widths, and branching ratios are obtained accordingly.

We discuss in the end the possible contribution of scalar top quarks in the $gg$ and $\gamma \gamma$ loops. We anticipate however that the eventual effect of scalar top quarks in the loop simply amounts to a redefinition of $c_t$, leaving the structure of the above equations unchanged\;\footnote{By using the same $c_t$ in the $c \bar c$ width, we introduce a slight inaccuracy in the branching ratios via the total width, the charm cross section being a very minor component of it.}. 

To find the best fit values of the $c$'s, we minimize the usual $\chi^2$  function:
\be
\chi^2=\sum_a\;\frac{\left[\mu_a -G_a(c_t,c_b,c_W)\right]^2}{\sigma_a^2}
\ee
where $\mu_a$ runs over the values of $\mu$, Eq. (\ref{muvalues}) given by ATLAS, CMS and Tevatron, $\sigma_a$ the corresponding error and $G_a$ the theoretical expressions introduced above in terms of the couplings. We have in all eleven experimental data and three parameters, namely eight degrees of freedom.

\begin{figure}[h!]
\begin{center}
\epsfig{height=7truecm, width=7truecm,figure=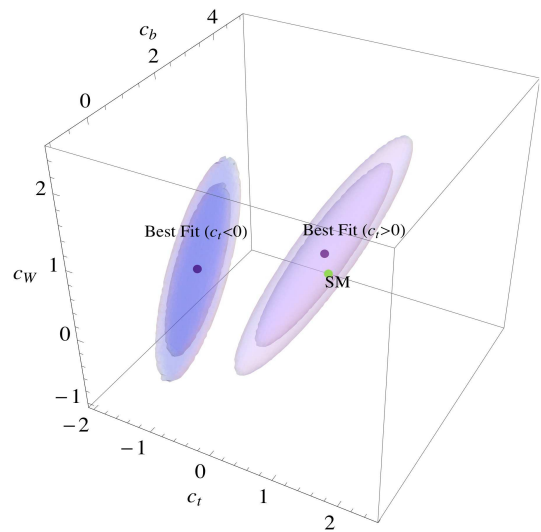}      
\caption{$1\sigma$ (inner) and $2\sigma$ (outer) regions in the space $(c_t,c_b,c_W)$ around the best fit points  obtained by minimizing the $\chi^2$ function. The fit includes ATLAS, CMS and Tevatron data, see text. Solutions are at $(\bar c_t, \bar c_b,\bar c_W)^+$=(0.886, 1.14, 1.19) and $(\bar c_t, \bar c_b,\bar c_W)^-$=(-0.908, 0.642, 0.666). 
We also add a point for the Standard Model value (green point).}
\label{fig:lobi}
\end{center}
\end{figure}

As a first step we identify the points in this space  which minimize the $\chi^2$ function. Setting conventionally $c_W \gtrsim 0$, we obtain the two solutions shown in Fig~\ref{fig:lobi}, corresponding to  $c_t<0$ and $c_t>0$. The two solutions have the same value of $\chi^2$, with $\chi^2/\mathrm{d.o.f}= 0.6$.

The relative sign of $c_t$ and $c_W$ is relevant to the $\gamma \gamma$ decay, see Eq. (\ref{2gamma}). It is no surprise that there is also a best fit solution with $c_t<0$, particularly in a situation where the central $\gamma \gamma$ signal exceeds the SM value~\cite{ATLAS,CMS}. A similar situation was also found in recent two-parameter fits~\cite{contino,ellis,grojean,riva}.

The inner and outer shells in Fig.~\ref{fig:lobi} represent the $1\sigma$ and $2\sigma$ regions. In the same figure we place the Standard Model point $(1,1,1)$, which falls within $1\sigma$ from the best fit to data.   We will focus on the positive solution. 

Next we study the surface  $S:(c_t(\tan\beta,m_H),c_b(\tan\beta,m_H),c_W(\tan\beta,m_H))$ predicted by Minimal Supersymmetry. Here we  briefly recall its definition. 

To start, let us recall that  in the basis ($H_d, H_u$), the mass matrix of the CP-even Higgs fields is given by:
\bea
&&{\cal M}_S^2=M_Z^2\left(\begin{array}{cc}  \cos^2\beta & -\cos\beta\sin\beta \\ -\cos\beta\sin\beta & \sin^2\beta \end{array}\right) 
+ M_A^2\left(\begin{array}{cc}  \sin^2\beta & -\cos\beta\sin\beta \\ -\cos\beta\sin\beta & \cos^2\beta \end{array}\right) +
\left(\begin{array}{cc} 0 & 0 \\0 &\delta \end{array}\right) 
\label{massmatrix}
\eea
with $\delta$ the radiative correction
\be
\delta = \frac{3 \sqrt{2}}{\pi^2\sin^2\beta } G_F (M_t)^4 \log\left(\frac{ \sqrt{M_{\tilde t_1}M_{\tilde t_2}}}{M_t}\right)
\label{stopcorr}
\ee
The first term in ${\cal M}_S^2$ arises from the so-called Fayet-Iliopoulos term determined by the gauge interaction. In the radiative corrections we have kept only the top-stop contribution. 
We determine the system of eigenvalues and eigenvectors of this matrix expressing $M_A$ in terms of the CP-even Higgs boson mass, $M_H$, and the known values for $M_Z$ and $M_h=125$~GeV, according to:
\be
M_A^2=M_H^2+M_h^2-M_Z^2-\delta
\label{trace}
\ee
The latter relation is obtained by equating the trace of ${\cal M}_S^2$ to $M_H^2+M_h^2$ (the sum of eigenvalues). 

In the calculation of eigenvalues and eigenvectors, we also express the quantity $t$, defined by: 
\be
t= \log\left(\frac{ \sqrt{M_{\tilde t_1}M_{\tilde t_2}}}{M_t}\right)
\ee
as a function of $\tan\beta$ and $M_H$, {\it i.e.} we find $t=F(\tan \beta, M_H)$.  This can be obtained from the eigenvalue difference squared, $(M_H^2-M_h^2)^2$. 
 The function $F$ arises as the solution of a second order equation. However, the other root is spurious, as it does not reproduce the degeneracy of the heavier Higgs boson masses as $M_H\to \infty$, see below, Eqs. (\ref{masses}). 
More on the function $F$ can be found in~\cite{Maiani:2012ij}. 

In the literature, it is customary to use $M_A$ as an independent parameter. Here, $M_A$ is obtained from eq. (\ref{trace}) and from the function $F(\tan \beta, M_H)$. In addition:
\be
M_{H^\pm}^2=M_A^2 +M_W^2
\ee

For future reference, we record some values of $M_A$ and $M_{H^\pm}$ (in GeV):
\begin{eqnarray}
&&M_A=[315,  358];\;  M_{H^\pm}=[325,  367]\;\;(M_H=360)  \nonumber \\
&&M_A=[600,  619];\;  M_{H^\pm}=[604,  624]\;\;(M_H=620) \nonumber \\
&& \tan\beta= [1,  6]
\label{masses}
\end{eqnarray} 

We next proceed to the computation of  the normalized eigenvector corresponding to $h(125)$, namely:
\be
{\bm S}_h(\tan\beta,M_H)=\begin{pmatrix}S_{hd}(\tan\beta,M_H)\\S_{hu}(\tan\beta,M_H)\end{pmatrix}
\ee
and define the surface $S$ parameterized by $\tan\beta$ and $M_H$ in the $(c_t,c_b,c_W)$ space:
\bea
&&c_t(\tan\beta,M_H)=\frac{\sqrt{1+\tan^2\beta}}{\tan\beta}\;S_{hu}(\tan\beta,M_H)\\
&&c_b(\tan\beta,M_H)=\sqrt{1+\tan^2\beta}\; S_{hd}(\tan\beta,M_H)\\
&&c_W(\tan\beta,M_H)=\frac{1}{\sqrt{1+\tan^2\beta}}\; S_{hd}(\tan\beta,M_H)+\frac{\tan\beta}{\sqrt{1+\tan^2\beta}}\; S_{hu}(\tan\beta,M_H)
\eea

\begin{figure}[h!]
\begin{center}
\epsfig{height=7truecm, width=7truecm,figure=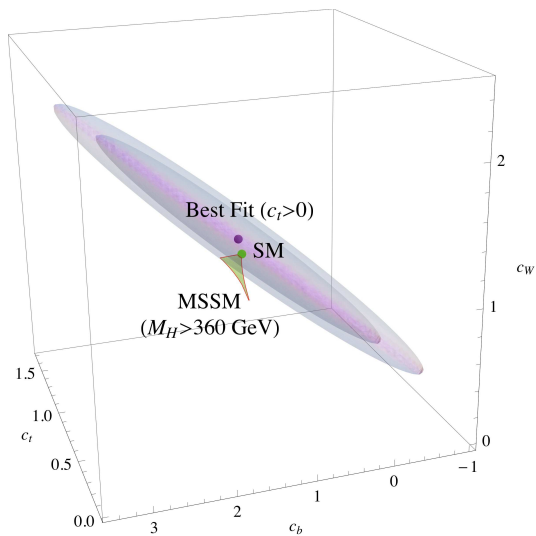}      
\caption{The intersection of the surface $S$ in the $(c_t,c_b,c_W)$ space, as predicted by MSSM, with the $1\sigma$ and $2\sigma$ regions around the best fit point to ATLAS and CMS data.}
\label{fig:lobivela}
\end{center}
\end{figure}

In Fig.~\ref{fig:lobivela} we show the $1\sigma$ and $2\sigma$ regions around the best fit solution with $c_t>0$,  intercepted by the surface $S$ described above, in the parameter range: 
\be
320\; {\rm GeV}<M_H<620\; {\rm GeV}; \;\;1<\tan\beta<6
\label{massrange}
\ee

The so called decoupling limit is manifested here in the fact that $S$ heads straight into the Standard Model point as the mass $M_H$ is increased. To be definite, we restrict to the mass range (\ref{massrange}).

\begin{figure}[h!]
\begin{center}
\epsfig{height=6truecm, width=8truecm,figure=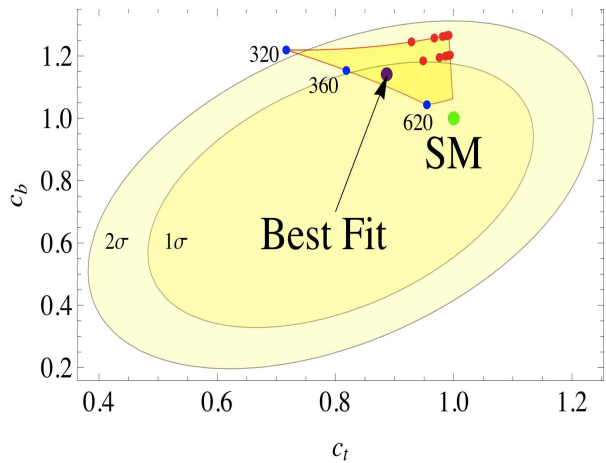}      
\caption{The surface $S$ as embedded in the $(c_t,c_b)$ space at $c_W=1$. The $1\sigma$ and $2\sigma$ regions are the intersections of the ellipsoids at $c_t>0$, Figs.~1 and~2 with the plane $c_W=1$. The outer side and the short inner side of $S$ correspond to $M_H=320$ and $M_H=620$ GeV, respectively. The intersection with the 1$\sigma$  border corresponds to $M_H= 360$ GeV. Red points mark values of $\tan\beta$ from 1 (lefthand side) to 6 (righthand side). The best fit point projects inside the MSSM surface, but is above it $(\bar c_W=1.19)$ in the three dimensional space.}
\label{fig:proj}
\end{center}
\end{figure}

A better display of the situation is obtained by fixing $c_W=1$  and considering the intersections of the $1\sigma,2\sigma$ ellipsoids with that plane. This is illustrated in Fig.~\ref{fig:proj} where we  also plot  the surface $S$ as embedded in the  $(c_t,c_b)$ space.

The (blue) points on the left side of $S$ correspond to $M_H=320$, $360$ and $620$~GeV, the intermediate value marking the border of the $1\sigma$ region.
 The (red) points crossing the surface from the left to the right are associated to different values of $\tan\beta$, from $1$, at the lefthand side, to $6$, at the righthand one.
 The best fit point projects inside the MSSM surface, but is above it $(\bar c_W=1.19)$ in the three dimensional space. 

The outcome of this analysis is that, even though we lost the footprints of an heavier Higgs at $320$~GeV, the hypothesis that the second $0^+$ particle predicted by Minimal Supersymmetry might just be lurking in the background is still viable. 

To 1$\sigma$, the region $M_H\gtrsim 360$~GeV is all allowed, along with $1<\tan\beta\lesssim 6$, consistently with the limits arising from the absence of Flavor Changing Neutral Current $B_s$ decays. Data are compatible with a regime of incomplete decoupling, the `intermediate - coupling regime' of Ref. \cite{Djuadi1}. However, MSSM could well be realized in the full decoupling mode.
Consistency to 1$\sigma$ with SM is also found in~\cite{Djuadi2}. 

In Ref.~\cite{Maiani:2012ij} we had reported the values of $\sigma \times BR$ for a hypothetical $H$ at $M_H=320$ and $500$ GeV. 
We close here by giving the picture for $M_H$ at $360$ and $620$ GeV and comment on production and decay of the corresponding pseudoscalar and charged Higgs particles, $A$ and $H^\pm$.

\begin{figure}[h!]
\begin{center}
\epsfig{height=6truecm, width=8truecm,figure=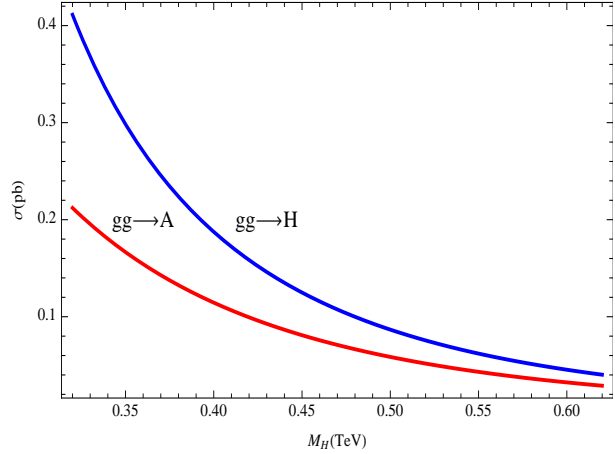}      
\caption{Production of $H$ and $A$ at the LHC (total energy 8 TeV) from gluon fusion. Cross sections (pb) are given for $\tan\beta =1.5$, as functions of $M_H$(TeV), in the $2\sigma$ allowed range. 
}
\label{xsect}
\end{center}
\end{figure}

Cross sections for $H$ and $A$ production by gluon fusion are shown in Fig. \ref{xsect} in the full 2$\sigma$ allowed mass range, $360<M_H<620 ~{\rm GeV}$. 
Cross sections have been computed with ALPGEN~\cite{alpgen}, and  corrected with the gluon K-factor taken from~\cite{Huston}.

\begin{figure}[h!]
\begin{center}
\epsfig{height=6truecm, width=8truecm,figure=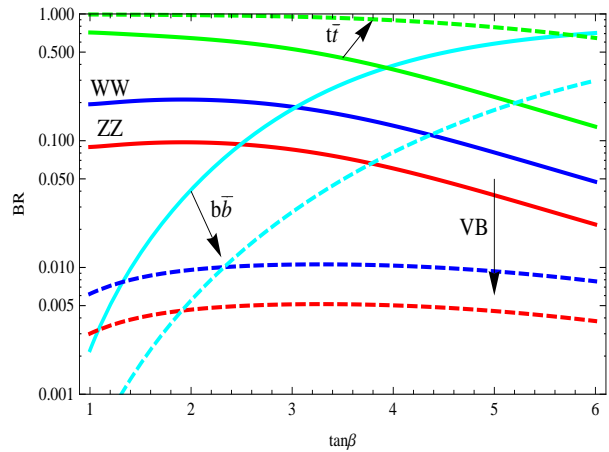}      
\caption{Branching ratios for the dominant deecay channels of $H$. Solid lines for $M_H= 360$ GeV, dashed ones for  $M_H= 620$ GeV. Decays are dominated by the  $t\bar t$ channel, whereas  vector boson and $b\bar b$ channels are still visible in the lower part of the mass range.
}
\label{bratio}
\end{center}
\end{figure}

Branching ratios for the dominant decay channels of $H$ are reported in Fig. \ref{bratio}, for $M_H= 360$ GeV, full lines, and $M_H= 620$ GeV, dashed lines, for $1<\tan\beta<6$. Decays are dominated by the  $t\bar t$ channel, with  vector boson and $b\bar b$ channels still visible in the lower part of the mass range.

$A$ and $H^\pm$ decay branching ratios can be obtained with  HDECAY~\cite{Hdecay}. In the $1\sigma$ allowed mass range, top quark channels, $t\bar t$ and $t\bar b$ respectively, are are by far the dominant ones.
Production at the LHC and decays of the $H^\pm$ have been extensively studied in Refs. \cite{chHprodc}. The dominant mechanism, in the mass range considered here, is by $b$-gluon fusion:
\be
b+g \to t+H^-
\ee
followed by $H^- \to \bar t +b$. 

Finally, we comment on possible effects of scalar top quarks in the $hgg$ and $h\gamma\gamma$ loops, in view of the large contribution of scalar tops to the Higgs mass correction, Eq. (\ref{stopcorr}). Such effects have been considered in \cite{dj,carena}.
Unlike the mass correction, the $\gamma\gamma$ and $gg$ loops are ultraviolet convergent, therefore heavy scalar particles decouple in the infinite mass limit. The only interesting case is the contribution of a light stop, the other one being correspondingly massive in order to satisfy Eq. (\ref{stopcorr}). This case has been considered in \cite{dj} and leads, neglecting mixing amongs scalar quarks,  to:
\be
\frac{A_{t+\tilde t}}{A_t} \approx 1+ \left(\frac{m_t}{M_{{\tilde t}_1}}\right)^2\frac{A_0(\tau)}{A_{1/2}(\tau_{t})}\simeq  1+ 0.25 \left(\frac{m_t}{M_{{\tilde t}_1}}\right)^2=f_t(M_{{\tilde t}_1})
\ee
where $\tau=4 M_{{\tilde t}_1}^2/m_h^2$ and $\tau_t=4 m_{t}^2/m_h^2$, and $A_{0,1/2}$ are defined in \cite{dj,carena}.
As noted before the additional contribution simply amounts to a redefinition of $c_t$ and therefore does not affect the best fit procedure. In the no-mixing approximation, the $\tan\beta$ dependency of the top and scalar top loop amplitudes is the same so that the effect of scalar top simply leads to an effective loop t-quark coupling: 
\be
c_t^{\rm loop}(\tan\beta,M_H)=f_t(M_{{\tilde t}_1})\; c_t(\tan\beta,M_H)
\ee
This amounts to a $3\%$ correction for  $M_{{\tilde t}_1}=500$ GeV.
 
In conclusion, present experimental information is consistent with the heavier scalar particles of MSSM to be in a relatively low energy region, between 320 and 620 GeV. A meaningful search can be done at the LHC, even at present luminosity.  We underscore that higher Higgs particles  still represent one of the most interesting test-tables of Minimal Supersymmetry. 



\bigskip 

\end{document}